\documentclass[journal]{IEEEtran}
\usepackage{amsmath,amsfonts}
\usepackage{algorithmic}
\usepackage{array}
\usepackage[caption=false,font=normalsize,labelfont=sf,textfont=sf]{subfig}
\usepackage{caption}
\usepackage{textcomp}
\usepackage{stfloats}
\usepackage{url}
\usepackage{xcolor}
\usepackage{verbatim}
\usepackage{graphicx}
\usepackage{epstopdf}
\usepackage{multirow}
\usepackage{booktabs}
\usepackage{longtable}
\usepackage{makecell}
\usepackage{tabularx}
\usepackage{xcolor}
\usepackage{float}
\usepackage{pifont}
\usepackage{hyperref}
\hypersetup{
    citecolor=green, % 文献引用的颜色设置为绿色
    linkcolor=blue, % 链接的颜色设置为蓝色
    urlcolor=blue % URL的颜色设置为蓝色
}

\def\BibTeX{{\rm B\kern-.05em{\sc i\kern-.025em b}\kern-.08em
    T\kern-.1667em\lower.7ex\hbox{E}\kern-.125emX}}
\usepackage{balance}
\begin{document}
\title{Towards Native AI in 6G Standardization: The Roadmap of Semantic Communication}
\IEEEspecialpapernotice{(Invited Paper) \vspace{-4mm}}
\author{ 
        Ping~Zhang,~\IEEEmembership{Fellow,~IEEE,}
        Xiaodong~Xu,~\IEEEmembership{Senior Member,~IEEE,}
        Mengying~Sun,~\IEEEmembership{Member,~IEEE,}
        Haixiao~Gao,
        Nan~Ma,~\IEEEmembership{Member,~IEEE,}
        Xiaoyun~Wang, 
        Ruichen~Zhang,~\IEEEmembership{Member,~IEEE,}
        Jiacheng~Wang,~\IEEEmembership{Member,~IEEE,}
        and Dusit~Niyato,~\IEEEmembership{Fellow,~IEEE}
        % \textit{(Invited Paper)}
\thanks{
Ping Zhang, Xiaodong Xu, Mengying Sun, Haixiao Gao, and Nan Ma are with the State Key Laboratory of Networking and Switching Technology, Beijing University of Posts and Telecommunications, Beijing 100876, China; 
Xiaoyun Wang is with the China Mobile Research Institute, Beijing 100053, China;
Ruichen Zhang, Jiacheng Wang, and Dusit Niyato are with the College of Computing and Data Science, Nanyang Technological University, Singapore. (\textit{Ping Zhang and Xiaodong Xu contributed equally to this work and should be considered co-first authors. Co-corresponding authors: Xiaodong Xu and Mengying Sun.} )
}
}

\maketitle
\begin{abstract}
Semantic communication (SemCom) has emerged as a transformative paradigm for future 6G networks, offering task-oriented and meaning-aware transmission that fundamentally redefines traditional bit-centric design. Recognized by leading standardization bodies including the institute of electrical and electronics engineers (IEEE) and the international telecommunication union (ITU), and actively discussed within the 3rd generation partnership project (3GPP) working groups, SemCom is rapidly gaining traction as a foundational enabler for native-AI 6G. This paper presents a comprehensive overview of recent progress in SemCom from both academic and industrial perspectives, with a focus on its ongoing and upcoming standardization activities. We systematically examine advances in representative application scenarios, architectural design, semantic-traditional system compatibility, unified evaluation metrics, and validation methodologies. Furthermore, we highlight several key enabling technologies, such as joint source-channel coding (JSCC), SemCom-based multiple access (MA) technologies such as model division MA (MDMA), and semantic knowledge base (KB), that support the practical implementation of SemCom in standard-compliant systems. Additionally, we present a case study for channel state information (CSI) feedback, illustrating the concrete performance gains of SemCom under 3GPP-compliant fading channels.  Finally, we discuss emerging challenges and research opportunities for incorporating semantic-native mechanisms into the evolving 6G standardization landscape, and provide forward-looking insights into its development and global adoption.
\end{abstract}


\section{Introduction}
As the global transition toward 6G accelerates, the integration of native-artificial intelligence (AI) with semantic communications (SemCom) heralds a paradigm shift: from traditional bit-centric architectures to meaning-aware, goal-oriented networking. Unlike previous generations, 6G is envisioned not merely as a faster communication system, but as a cognitively intelligent infrastructure capable of understanding, reasoning, and interacting with both users and devices. Realizing this vision requires the deep embedding of native-AI across all layers of the communication system, from the physical layer to application-level interfaces. 

In the evolution of 6G standardization, the design of native-AI aims to natively integrate AI/machine learning (ML) frameworks into the network architecture, establishing an intelligence-driven system for automation, optimization, and efficient network operation and management. It emphasizes full life-cycle management (LCM) of AI models, including configuration, monitoring, evolution, and transition, and enables AI-as-a-Service (AIaaS) capabilities within the core network \cite{guo2024survey}. An extensible AI/ML framework, evolved from 5G-Advanced, is constructed to support dynamic distribution and updates of models, data, and knowledge. A novel ``data plane'' is introduced to facilitate SemCom and intelligent task deployment, while both the control and user planes are made intent-programmable, allowing AI to map user intent to multidimensional network resources and enable differentiated services. SemCom plays a pivotal role by enhancing the transmission efficiency of task-relevant information, and supporting AI agent collaboration, knowledge sharing, and policy optimization through semantic quality of service (QoS) and intent recognition, empowering 6G to evolve toward a new semantic-driven and native-AI communication paradigm \cite{zhang2024intellicise, yang2022semantic}.

\subsection{Challenges}
However, the standardization of SemCom for native-AI 6G networks presents a series of complex and multi-dimensional challenges. This work identifies and analyzes the critical hurdles that must be addressed to establish a unified, scalable, and intelligent framework for SemCom toward 6G. The major challenges are outlined as follows.

\begin{itemize}
    \item \textbf{Typical Application Scenarios and Unified Evaluation Metrics:} SemCom enables demand-oriented transmission, but its effectiveness and evaluation metrics vary across application domains. 6G heterogeneous scenarios, such as AI agent, immersive communications, satellite communications, and generative AI, exhibit diverse requirements and evaluation needs~\cite{3gpp_TR22_870_v030_2025}. Traditional metrics such as throughput and bit error rate fail to reflect semantic effectiveness. Thus, novel cross-layer metrics such as task accuracy, semantic fidelity, and intent alignment accuracy, must be formalized and benchmarked across representative scenarios to ensure fair comparison and standard compliance.
    \item \textbf{Compatibility with Existing Communication Systems:} Integrating SemCom with 6G requires maintaining backward compatibility. Unlike traditional syntax-based communication systems, SemCom relies on semantic extraction and task-driven AI-based coding and modulation, thereby challenging legacy design principles~\cite{guo2024survey, zhang2024intellicise}. The resulting compatibility issues stem from multiple sources: the mismatch between bit-level reliability metrics and task-oriented semantic performance goals, the difficulty of integrating semantic encoders/decoders into fixed protocol layering, the need to redesign signaling and hybrid automatic repeat request (HARQ) procedures to support semantics-aware processing, and the computational heterogeneity across devices that complicates large-scale deployment.
    \item \textbf{Generalized Design Framework:} Current SemCom designs are task-specific, limiting generalization. Standardization requires unified, modular architectures that abstract core components such as semantic encoders/decoders, while ensuring interoperability and extensibility across heterogeneous devices. A key challenge is designing efficient, general-purpose interfaces and protocols adaptable to diverse use cases without compromising performance.
    \item \textbf{Challenges in Supporting Multiple Devices:} Enabling SemCom on heterogeneous devices requires adaptive encoding, model alignment, and knowledge synchronization to address variations in AI processing and support multi-device coordination. Lightweight models are essential for resource-constrained devices \cite{zhang2024intellicise}, while continuous local data collection ensures on-device model updates tailored to specific communication tasks.
\end{itemize}

Despite its promising gains, SemCom still lacks the standard foundations needed for practical deployment in native-AI 6G networks. Current solutions are largely task-specific, with semantic metrics that are hard to benchmark across heterogeneous scenarios, devices, and layers. Meanwhile, embedding semantics-aware coding, resource allocation, and knowledge-assisted processing into legacy protocol stacks introduces major interoperability and signaling challenges. These gaps motivate this paper to develop a unified, scalable, and backward-compatible SemCom framework from a standardization-driven perspective.

\subsection{Related Works}
In the 6G era, the convergence of communications and AI is driven by the transformation of networks from mere connectivity pipes into intelligent information platforms. SemCom enables intent-aware feature extraction, significantly reducing resource usage while ensuring service quality, embodying the 6G native-AI paradigm of integrated perception, understanding, transmission, and service. Researchers from both industry and academia have been striving to realize this vision, and standardization organizations such as 3GPP, the institute of electrical and electronics engineers (IEEE), and the international telecommunication union (ITU) are actively exploring pathways to make it a reality. 

ITU has made significant progress in the research and standardization of SemCom across multiple study groups. In SG13, SemCom has been identified as a key enabler for future networks beyond IMT-2020. This includes the development of technical reports and recommendations such as \textit{TR.Reqts-SAN} and \textit{Y.RA-SAN}, which defines the requirements and reference architectures for semantic-aware networking. Additionally, SemCom scenarios integrated with decentralized technologies have been studied under \textit{TR.SIC-DLT}. In SG16, the focus has expanded to semantic processing of unstructured multimedia data, leading to specifications such as \textit{F.JSQSUDA} for joint semantic query systems. Meanwhile, in 2021, the SG20 of ITU-T initiated the \textit{YSTR.SemComm.IoT} project titled ``Architectural Framework for SemCom Services for Internet of Things (IoT) and Smart City \& Community.''  Recently, led by Pengcheng Laboratory, in collaboration with Beijing University of Posts and Telecommunications (BUPT) and China Telecom, a new ITU-T technical report (TR) entitled ``Architectural Framework for Knowledge-Based Semantic Communication over Public IMT Network'' was initiated\hyperlink{footnote1}{\footnotemark[1]}. This milestone signifies that SemCom has progressed from conceptual research to the stage of international standardization, providing a unified framework and roadmap to guide subsequent system design and industrial deployment. 

\hypertarget{footnote1}{\footnotetext[1]{ITU-T YSTR.Af-KBSC,  PCL, BUPT, China Telecom, ``Architectural Framework for Knowledge Based Semantic Communication over Public IMT Networks,'' Available: \url{https://www.itu.int/ITU-T/workprog/wp_item.aspx?isn=23408}}}

Academically, SemCom can be traced back to Shannon and Weaver's three-level communication model, which comprises the syntactic, semantic, and pragmatic levels. Driven by rapid AI advances, SemCom has re-emerged as a key focus in next-generation communication research. Advancing 6G Standardization in 3GPP, the transformative potential of SemCom and its critical role in enabling 6G standardization are increasingly recognized by the industry. These efforts demonstrate the commitment of industry, academia, and standards bodies to embedding semantic awareness into the next generation of native-AI communication systems, and the related efforts are summarized in Table~\ref{tab:related works}. 

%Table~\ref{tab:related works} summarizes representative industry and academic SemCom directions. Industry efforts emphasize compatibility-oriented design (e.g., semantic-level HARQ), token-based communication, and AI-assisted JSCC/JSCM for CSI compression and feedback. Academic studies mainly investigate SemCom techniques for multimodal data (e.g., images, videos, and speech), SemCom-enabled MA, and semantic KB for SemCom.

\begin{table*}[t]
    \renewcommand{\arraystretch}{1.1}
    \centering
    \begin{tabular}{|>{\centering\arraybackslash}m{2cm}|>{\centering\arraybackslash}m{3.1cm}|m{11cm}|}
    \hline 
    
    \multicolumn{3}{|c|}{\textbf{Overview of Industry SemCom Solutions}\hyperlink{footnote2}{\footnotemark[2]}} \\
    
    \hline
    \textbf{\textit{Company}} & 
    \textbf{\textit{Focus Direction}} & 
    \textbf{\textit{Representative Solution}} \\
    
    \hline
    China Mobile Communication Corporation (CMCC) 
    &  Compatibility design of SemCom and semantic-level HARQ
    &  CMCC has explored the compatibility design of the SemCom architecture with cross-layer communication systems and investigated verification and HARQ mechanisms, enabling transmission tailored to differentiated error sensitivity. \\

    \hline
    
    Huawei
    & Split inference and Token communication
    & Huawei focuses on that distributed split inference, together with token‑based communication, enhances mobile AI services by reducing latency and boosting user satisfaction through reliable and adaptive interactions. \\

    \hline
    Zhongxing Telecommunication Equipment Corporation (ZTE)
    & JSCC/JSCCM for CSI, AI-based coding and modulation
    & ZTE focuses on AI-based JSCC and joint source-channel coding and modulation (JSCCM) for CSI compression and feedback, proposes AI-enhanced coding and modulation schemes, and introduces intent-driven services enabled by AI. \\

    \hline

    Pengcheng Laboratory
    & SemCom and KB-enabled SemCom 
    & Pengcheng Laboratory highlights that SemCom, powered by JSCC, increases spectral efficiency, while a semantic KB makes the network more intelligent, highly reliable, and low‑latency. \\
    \hline
    
    Lenovo
    & SemCom for CSI and AI-based receiver design
    & Lenovo emphasizes native-AI 6G networking, prioritizing two-sided models based on JSCC for AI-driven symbol generation and detection. These models also enable efficient CSI compression and support AI-based receiver design by integrating channel estimation, equalization, demodulation, and symbol reconstruction into a unified framework. \\

    \hline
    
    Qualcomm
    & JSCC for control (e.g., CSI and HARQ)
    & Qualcomm emphasizes AI-enabled 6G air interface by adopting JSCC for air interface control (e.g., CSI enhancement), enhances the LCM, and further explores customized models and extends AI/ML-embedded modules across broader domains. \\

    \hline
    Lucky GoldStar Corporation (LG)
    & SemCom and AI processing
    & For 6G communication-computation convergence, LG identifies SemCom and AI-based source coding as potential enabling technologies. \\

    \hline
     
    Korea Telecom (KT)
    & SemCom
    & KT emphasizes that SemCom is regarded as one of 6G key enablers. \\
    
    \hline

    \multicolumn{3}{|c|}{\textbf{Overview of Academic SemCom Research}} \\

    \hline
    
    \textbf{\textit{Direction}} & 
    \textbf{\textit{Representative Paper}} & 
    \textbf{\textit{Technical Features}} \\
    
    \hline
    
    \multirow{4}{*}[-5.5mm]{JSCC} 
    & Deep JSCC-based image transmission \cite{xu2021wireless}
    & This method uses channel‑wise soft attention to dynamically rescale encoded features according to the real-time signal-to-noise ratio (SNR), thus balancing source compression with robustness. \\
    
    \cline{2-3}
     
    & Deep JSCC-based video transmission \cite{tung2022deepwive}
    & This framework employs the deep neural network to map videos directly to over‑the‑air symbols, and uses reinforcement learning (RL) to dynamically optimize channel bandwidth resources. \\

    \cline{2-3}
    
    & \makecell{Iterative JSCC \\ text transmission \cite{yao2022semantic}}
    & This semi‑neural scheme iteratively feeds residual semantics of decoded text to the next round, guiding soft semantic synthesis at the receiver. \\

    \cline{2-3}

    & Nonlinear transform-based deep speech transmission \cite{xiao2023wireless}
    & This method generates the over-the-air speech sequence through a nonlinear transform combined with JSCC and employs an entropy model to assess feature importance and allocate coding rates. \\

    \hline

    \multirow{3}{*}[-2mm]{\makecell{SemCom-\\based MA}}
    
    & RSMA‑based panoramic video semantic transmission \cite{gao2025rate}
    & This solution applies rate-splitting multiple access (RSMA) to adaptively split semantic streams based on field-of-view, and jointly optimizes power, common rate, and bandwidth allocation. \\ 
    
    \cline{2-3}
    
    & MDMA \cite{zhang2023model}
    & This scheme extracts common and private semantics in the model domain to enable orthogonality and concurrent transmission of multi-user signals, extending JSCC’s robustness to the capacity. \\

    \hline

    \multirow{2}{*}[-0.7mm]{\makecell{SemCom-\\based CSI \\ Feedback}}
    & \makecell{Deep JSCC‑based \\ CSI feedback \cite{xu2022deep}}
    & This framework learns CSI representations and performs compression in an end-to-end manner, while incorporating an SNR-adaptive scheme to cope with channel variations. \\

    \cline{2-3}

    & Joint coding-modulation CSI feedback \cite{ren2025semcsinet}
    & This scheme integrates the channel quality indicator into the CSI feedback and employs a joint coding and modulation scheme to digitally convey CSI over noisy feedback channels. \\
    
    \hline

    \multirow{2}{*}[-2mm]{\makecell{Semantic \\ KB}}
    & Codebook-based hierarchical semantic KB \cite{wang2024unified}
    & This approach enables efficient and scalable KB support for multitask SemCom by laterally sharing and vertically aggregating semantic subspaces. \\

    \cline{2-3}

    & KB-based residual text SemCom \cite{yi2023deep}
    & This framework matches each message in a shared KB, transmits only the residual data, and fuses it with similar features at the receiver to reconstruct the information. \\

    \hline
    \end{tabular}
    \caption{Overview of industry and academic progress in SemCom. The upper section lists representative solutions from companies such as LG and ZTE, highlighting how the industry is advancing native-AI networks and knowledge‑driven architectures. The lower section classifies seminal academic papers under four research directions, i.e., JSCC, SemCom-based MA, SemCom-based CSI feedback, and semantic KB, which summarizes their technical features to serve as a reference for standardization and future technological evolution.}
    \label{tab:related works}
\end{table*}

\begin{figure*}[!t]
    \centering
    \includegraphics*[width=170mm]{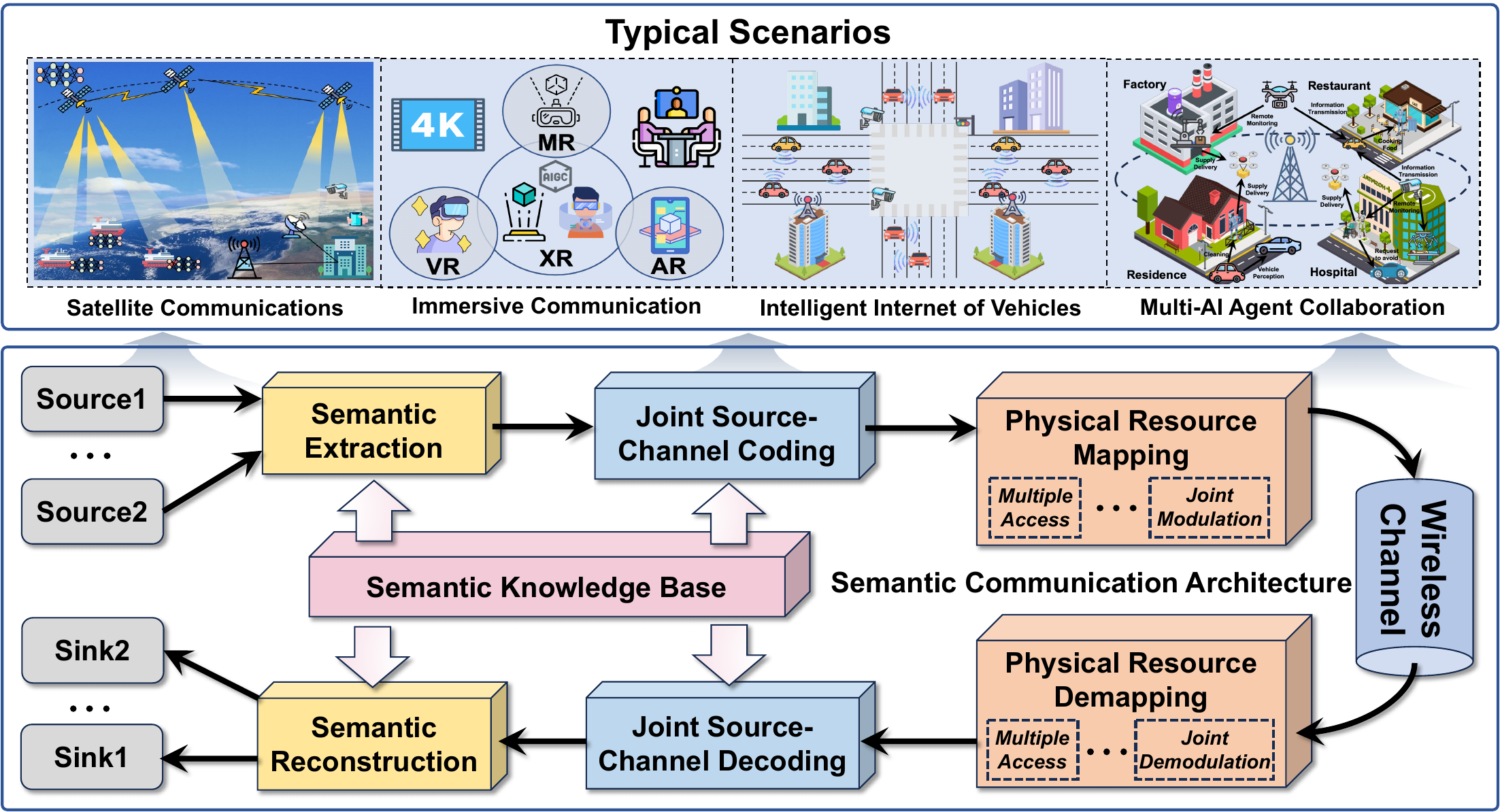}
    \caption{Typical scenarios and E2E architecture of 6G SemCom. The upper part of the illustration shows four emblematic use cases, i.e., satellite communications, immersive communication, intelligent internet of vehicles, and multi-AI agent collaboration. The lower part depicts a SemCom architecture that underpins them. By supporting these scenarios, the SemCom architecture enables optimized spectrum efficiency, reduced end-to-end latency, and context-adaptive services across heterogeneous network layers and application domains.}
    \label{fig:Scenario_SemCom}
\end{figure*}

\subsection{Contributions}

To strengthen the research directions, we focus on several key issues in advancing SemCom standardization. First, it is necessary to clarify how SemCom capabilities can be integrated into the 6G system architecture to form a unified SemCom system. Second, compatibility between semantic and traditional communication systems must be achieved to ensure gradual deployment and interoperability. Third, key enabling technologies are explored for supporting the SemCom standardization. Finally, a unified platform and verification mechanism are essential for standardization, requiring a cross-vendor, cross-application, and cross-network test framework for SemCom. Overall, the contributions of this article are summarized as follows.
\begin{itemize}
    \item \textbf{Architecture and Compatibility Design:} We propose SemCom framework for native-AI 6G networks and design a compatibility mechanism that ensures seamless coexistence between semantic and traditional communication systems. 
    % The framework supports the progressive introduction of semantic functions and their co-evolution with existing networks.
    
    \item \textbf{Typical Scenarios, Evaluation Metrics, and Key Enabling Technologies:} We present representative application scenarios for SemCom and build a comprehensive evaluation metric framework that merges traditional and semantic indicators. We further identify and analyze key enabling technologies including joint source-channel coding (JSCC), SemCom-based multiple access (MA) such as model division MA (MDMA), SemCom-based channel state information (CSI) feedback, and semantic knowledge base (KB) technologies.
    
    \item \textbf{Unified Implementation and Verification Method:} We propose a cross-layer verification framework for SemCom, integrating semantic-aware QoS mapping and cross-layer design principles, thereby enabling verifiable and reproducible evaluation to support standardization.
    
    \item \textbf{Path Forward for SemCom Standardization:} We validate the performance benefits of SemCom through simulation of CSI feedback. Based on this, we outline a standardization roadmap that evolves from control-plane enablement to data and user plane integration, providing a structured path for SemCom standardization.
    
\end{itemize}

\hypertarget{footnote2}{\footnotetext[2]{Company slide decks from the 3GPP workshop on 6G (Incheon, March 10-11, 2025) can be downloaded at \url{https://www.3gpp.org/ftp/workshop/2025-03-10_3GPP_6G_WS/Docs/}. The specific archives for various companies are \textit{Huawei:} 6GWS‑250159.zip, \textit{ZTE:} 6GWS‑250008.zip, \textit{Lenovo:} 6GWS‑250061.zip, \textit{Qualcomm:} 6GWS‑250068.zip, \textit{LG:} 6GWS‑250024.zip, \textit{Pengcheng Laboratory:} 6GWS‑250194.zip, and \textit{KT:} 6GWS‑250020.zip.}}

\section{Typical Scenarios and Evaluation Metrics}

This section introduces four representative SemCom-enabled scenarios and a multi-dimensional evaluation framework that integrates semantic-level metrics with conventional communication metrics to enable cross-layer validation.
\subsection{Typical Scenarios}

As illustrated in Fig.~\ref{fig:Scenario_SemCom}, this section provides an overview of several representative application scenarios where SemCom is expected to play a pivotal role in future 6G networks. 

\subsubsection{Satellite Communications}
In satellite communication scenarios, three key applications emerge: remote sensing image downlink, voice call, and satellite internet data services \cite{3gpp_TR22_870_v030_2025}. However, transmissions in satellite scenarios are constrained by highly dynamic, resource-limited, and severely fading satellite–ground channels. SemCom addresses these challenges by JSCC and enabling semantic compression of images, videos, audio, and task features, thereby reducing transmission redundancy and lowering bandwidth consumption. For instance, in response to high-data-rate requests, the system can perform real-time semantic analysis to extract critical information, such as key infrastructure elements, or emergency risk events, from videos and images. This ensures that task-relevant content can be preserved even under ultra-high compression, making it well suited to the resource-constrained transmission conditions in satellite communications.

\subsubsection{Immersive Communication}

Immersive communications is one of the six key scenarios defined by ITU-R in the IMT-2030 framework. The scenario encompasses XR remote collaboration, lifelike holographic meetings, and multisensory telepresence, and it requires sub-millisecond time-synchronized delivery of 3D video, spatial audio, haptic feedback, and environmental parameters \cite{gao2025rate}. Compared with traditional video services, the system must support scalable downlink throughputs ranging from hundreds of megabits to gigabits per second and achieve end-to-end (E2E) latency below five milliseconds while maintaining high reliability and ultra-low jitter to prevent motion sickness and preserve immersion. For such scenarios, SemCom can prioritize perceptually salient and task-critical semantics (e.g., objects of interest, user interactions, and scene dynamics), or transmit a low-resolution video to enable adaptive rendering and content reconstruction at the receiver. This facilitates low-latency, high-fidelity immersive experiences over bandwidth-limited and time-varying wireless channels, thereby making large-scale immersive services feasible in future 6G networks.

% By extracting and conveying task-level semantic features in place of redundant bit streams, SemCom enables meaning-first traffic optimization that substantially reduces transmission overhead without compromising the quality of experience (QoE) of users.

\subsubsection{Intelligent Internet of Vehicles}
The intelligent internet of vehicles enables real-time collaboration among vehicles,  overcoming the line-of-sight limits of individual sensors and giving autonomous driving and smart traffic global perception and prediction capabilities. A related use case presented in 3GPP TSG‑SA WG1\hyperlink{footnote3}{\footnotemark[3]} describes vehicles that detect areas outside their own sensing range and request cooperative perception data through vehicle-to-everything (V2X) networks. The base station authenticates the request, selects cooperative vehicles, and orchestrates semantic-level data sharing, where AI-encoded perception features are transmitted, fused, and decoded to generate a complete, real-time environmental map, all while ensuring communication reliability, dynamic adaptation, and data security.

\hypertarget{footnote3}{\footnotetext[3]{3GPP TSG‑SA WG1 Meeting \#110, ``Use case on AI‑driven multi‑vehicle cooperative perception,'' S1‑252967, May 2025. Available: \url{https://www.3gpp.org/ftp/tsg_sa/WG1_Serv/TSGS1_110_Fukuoka/Docs/S1-252967.zip}}}

\subsubsection{Multi-AI Agent Collaboration} 
As AI agents, empowered by large language models, on-device inference, and edge-AI services, proliferate across domains such as productivity, logistics, and smart environments, the demand for task-oriented, intent-aware communication is growing rapidly~\cite{3gpp_TR22_870_v030_2025}. SemCom supports this by allowing agents to exchange high-level intents, contextual knowledge, and semantic tokens, rather than raw data. In 6G-enabled multi-agent systems, agents can dynamically form sub-networks to decompose tasks, synchronize local semantic KBs, and coordinate actions in real time. 
Supported by SemCom and edge model adaptation, the agents transmit only task-relevant semantic features, reducing overhead and enabling them to fulfill complex user-defined goals in a scalable and context-aware manner. In a nutshell, SemCom empowers multi-agent collaboration through intent-level exchange, semantic-aware encoding/scheduling, knowledge alignment, and communication-efficient cooperative decision-making, improving scalability, efficiency, and robustness.

\begin{table*}[t]
    \renewcommand{\arraystretch}{1.1}
    \centering
    \begin{tabular}{|>{\centering\arraybackslash}m{3cm}|>{\centering\arraybackslash}m{4.8cm}|>{\arraybackslash}m{8.8cm}|}
    \hline 
    
    \textbf{\textit{Category}} & 
    \textbf{\textit{Metrics}} & 
    \textbf{\textit{Descriptions}} \\

    \hline
    \textbf{Unified Evaluation} & 
    Semantic Similarity (SS) & 
    Cosine similarity between semantic vectors obtained after semantic extraction. \\

    \hline

    \multirow{3}{*}{\vspace{-15pt} \textbf{Image / Video}} &
    \makecell{Peak Signal-to-Noise \\ Ratio (PSNR) \cite{yang2022semantic,xu2021wireless,tung2022deepwive}} &  
    Log-scale measure of pixel differences between the reconstructed image (or frame) and the original one. \\

    \cline{2-3}
    
    &
    \makecell{Structural Similarity \\ Index Measure (SSIM) \cite{yang2022semantic}} &  
    Assesses image (or frame) similarity in terms of luminance, contrast, and structure. \\

    \cline{2-3}
    
    &
    Multi-Scale Structural Similarity Index Measure (MS-SSIM) \cite{tung2022deepwive}&  
    Computes and weights SSIM across multiple resolutions. \\

    \hline

    \multirow{4}{*}{\vspace{-35pt} \textbf{Text}} &
    Word Error Rate (WER) \cite{yao2022semantic} &  
    Percentage of insertions, deletions, and substitutions in the recognized text relative to the reference word count. \\

    \cline{2-3}
    
    &
    Similarity \cite{yang2022semantic,yi2023deep} &  
    Maps texts to vectors and calculates cosine similarity to measure literal semantic similarity. \\

    \cline{2-3}
    
    &
    Bidirectional Encoder Representation from Transformers (BERT) score \cite{yang2022semantic} &  
    Uses BERT embeddings to compute token-level semantic similarity between a candidate sentence and a reference sentence. \\

    \cline{2-3}
    
    &
    Bilingual Evaluation Understudy (BLEU) score \cite{yang2022semantic,yao2022semantic,yi2023deep} &  
    Evaluates machine-translated/generated text quality based on n-gram matches with length penalty. \\

    \hline

    \multirow{3}{*}{\vspace{-10pt} \textbf{Speech}} &
    Signal-to-Distortion Ratio (SDR) \cite{yang2022semantic} &  
    Ratio of reconstructed speech signal energy to distortion energy. \\

    \cline{2-3}
    
    &
    Perceptual Evaluation of Speech Quality (PESQ) \cite{yang2022semantic,xiao2023wireless} &  
    Simulates auditory perception and aligns reference and distorted speech signals, maps their perceptual difference onto a predicted MOS quality score.  \\

    \cline{2-3}
    
    &
    Mel Cepstral Distortion (MCD) &  
    Measures distortion of speech via differences in Mel cepstral coefficients. \\

    \hline

    \multirow{3}{*}{\makecell{\textbf{Classification \&} \\ \textbf{Object Detection}}} &
    Accuracy \cite{wang2024unified} &  
    Proportion of correctly classified samples out of all samples. \\

    \cline{2-3}
    
    &
    Precision &  
    Proportion of true positives among samples predicted as positive. \\

    \cline{2-3}
    
    &
    Recall &  
    Proportion of true positives detected among all actual positive samples. \\

    \hline

    \multirow{2}{*}{\makecell{\textbf{Semantic \&} \\ \textbf{Instance Segmentation}}} &
    Intersection over Union (IoU) &  
    Area of overlap between the predicted region and the ground truth divided by their union area. \\

    \cline{2-3}
    
    &
    Mean IoU (MIoU) \cite{wang2024unified} &  
    Average IoU across all classes. \\

    \hline

    \multirow{4}{*}{\vspace{-10pt} \textbf{Image Retrieval}} &
    Mean Average Precision (MAP) &  
    Mean of Average Precision (AP) values across all queries. \\

    \cline{2-3}
    
    &
    mAP@k &  
    MAP calculated using only the top-k results for each query. \\

    \cline{2-3}
    
    &
    \makecell{Normalized Discounted \\ Cumulative Gain (nDCG)} &  
    Rank-discounted cumulative relevance gain, normalized by the ideal ranking. \\

    \cline{2-3}
    
    &
    Recall@k &  
    Proportion of relevant items found in the top-k retrieval results. \\
    
    \hline
    
    \multirow{2}{*}{\makecell{\textbf{SemCom} \\ \textbf{Transmission Efficiency}}} &
    % Semantic Feasibility (SF) &  
    % Ratio of the source-coding rate to the channel’s available transmission rate. \\

    % \cline{2-3}
    
    Channel Bandwidth Ratio (CBR) \cite{tung2022deepwive} &  
    Ratio of the symbol numbers output by JSCC to the original symbol numbers. \\

    \cline{2-3}
    
    &
    Semantic Spectral Efficiency \cite{guo2024survey} &  
    Amount of effective semantic information conveyed per unit bandwidth. \\

    \hline
    \end{tabular}
    \caption{Comprehensive evaluation metrics for 6G SemCom. The table groups representative metrics by data modality (image, text, video, speech), core AI tasks (classification, object detection, semantic segmentation, image retrieval), and SemCom transmission efficiency (CBR, semantic spectral efficiency), enabling systematic comparison of source-channel designs and E2E performance.}
    \label{tab:evaluation metrics}
\end{table*}

\subsection{Evaluation Metrics}
We establish a comprehensive SemCom evaluation framework based on different modalities and diverse tasks, as shown in Table~\ref{tab:evaluation metrics}. This framework not only incorporates typical reconstruction metrics for multimodal data such as images \cite{xu2021wireless}, text \cite{yao2022semantic, yi2023deep}, video \cite{tung2022deepwive, gao2025rate} and speech \cite{xiao2023wireless}, but also includes evaluation criteria for downstream tasks such as classification, object detection, semantic segmentation and image retrieval \cite{wang2024unified}. More importantly, we propose a unified metric named semantic similarity (SS) that can be applied across modalities, which can also serve as a cross-modal evaluation metric for 3GPP use case\hyperlink{footnote4}{\footnotemark[4]}. For each modality, the original information is first mapped into a shared semantic space by a pretrained model, for example, a Vision Transformer for images and BERT for text, and then the cosine similarity between the resulting embeddings is calculated to measure cross-modal consistency. This unified demand-oriented metric framework provides a reliable basis for objectively evaluating SemCom performance in multimodal and multi-task scenarios.
Specifically, modality alignment is achieved by training modality-specific encoders with a shared objective, typically via contrastive learning or task-driven supervision, to project heterogeneous modalities into a unified semantic embedding space.
Representative models include CLIP and ALIGN for image-text alignment, AudioCLIP for audio-image-text alignment, and VideoCLIP/Florence for video-language alignment, which map semantically equivalent cross-modal content to highly similar representations.

\hypertarget{footnote4}{\footnotetext[4]{For example, SS can be used to assess text-video semantic similarity in 3GPP TSG-SA WG1 Meeting \#110, ``New use case on AI text-to-video generation supported by computing,'' S1-252949, May 2025. Available: \url{https://www.3gpp.org/ftp/tsg_sa/WG1_Serv/TSGS1_110_Fukuoka/Docs/S1-252949.zip}}}

In addition to semantic-level metrics such as peak signal-to-noise ratio (PSNR) and word error rate (WER), which assess reconstruction quality, semantic fidelity, and linguistic accuracy, traditional communication metrics are also incorporated into the evaluation framework to enable comprehensive validation. Conventional metrics such as spectral efficiency (SE), latency, and reliability reflect physical and transport layer performance, supporting both the evaluation of SemCom transmission efficiency and the development of unified cross-layer frameworks. Furthermore, integrating traditional and semantic metrics such as channel bandwidth ratio (CBR), semantic SE, and noise robustness enables the flexible construction of multi-dimensional evaluation metrics tailored to diverse requirements~\cite{guo2024survey, yang2022semantic,tung2022deepwive}. 

%Collectively, these representative scenarios and their multi-dimensional metrics, ranging from semantic similarity and perceptual quality (e.g., SS/PSNR/SSIM), to language/speech intelligibility (e.g., WER/BERT/BLEU/MCD), to task-level performance (e.g., accuracy/IoU/mAP), and transmission efficiency (e.g., CBR), highlight that SemCom must be evaluated beyond bit-level reliability. 
%Therefore, it is necessary to establish a unified and reusable evaluation metric system and benchmarking methodology for SemCom, tailored to different scenarios and application requirements. This will provide critical foundational support for comparability verification, standardization adoption, and large-scale deployment of diversified 6G services.
% Collectively, these scenarios and multi‑dimensional evaluation metrics indicate the need for a unified SemCom architecture and its corresponding enabling technologies to translate these requirements into deployable 6G systems.

\section{Semantic Architecture and Key Enabling Technologies}
This section is organized into two parts. It first presents the SemCom architecture for native-AI 6G, highlighting the key functional blocks and their interactions. It then introduces four enabling technologies such as JSCC, SemCom-enabled MA, SemCom-based CSI feedback, and semantic KB.

\subsection{Semantic Communication Architecture}

As shown at the bottom of Fig.~\ref{fig:Scenario_SemCom}, we propose an E2E SemCom architecture tailored for native-AI 6G networks. At the transmitter, semantic extraction first distills task-relevant features from raw multimodal inputs, and these features are then processed by a JSCC module that combines compression and error protection to produce robust latent representations. Both stages draw on a shared KB to supplement incomplete context information. The resulting latent vectors are subsequently mapped to physical-layer resources (e.g., transmit power and wireless channels) according to their semantic importance, enabling resilient transmission under dynamic channel conditions. At the receiver, the process is reversed through resource demapping, knowledge-assisted JSCC decoding, and semantic reconstruction to recover the original task-relevant information for downstream applications. By integrating semantic awareness, adaptive coding and modulation, and KB-guided reconstruction, this architecture achieves low-overhead, low-latency, and highly-reliable communication, providing a scalable and practical framework for the deployment of SemCom in future 6G networks.

\begin{figure*}[t]
    \centering
    \includegraphics*[width=180mm]{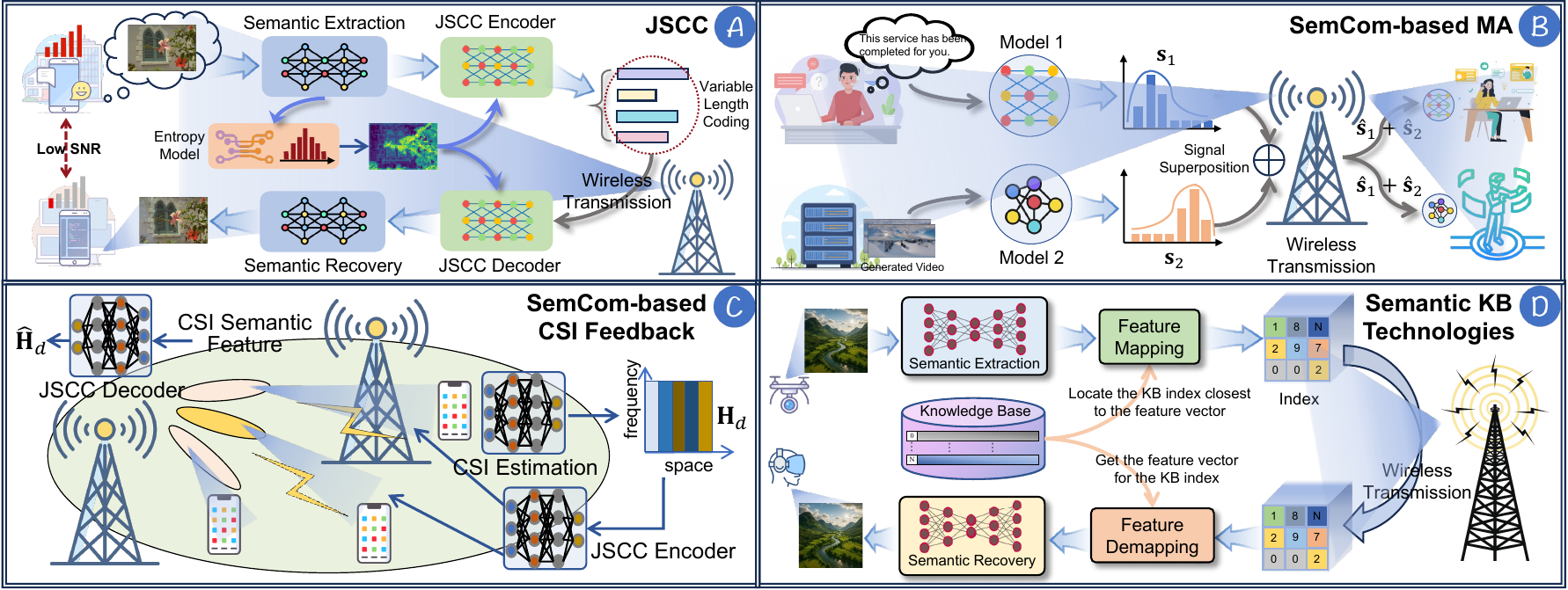}
    \caption{Illustrative framework of key enabling SemCom technologies for 6G. The diagram visualizes the end‑to‑end data flow, functional interplay, and evolution pathway of four core technologies, i.e., A: JSCC, B: SemCom-based MA, C: SemCom‑based CSI Feedback, and D: Semantic KB Technologies.}
    \label{fig:SemCom_Key_Technologies}
\end{figure*}

\subsection{Key Enabling Technologies}
Building on the SemCom architecture, the realization of native-AI 6G requires coordinated advances in key enabling technologies. Fig.~\ref{fig:SemCom_Key_Technologies} illustrates four enablers that provide a unified design framework for future specifications and drive the evolution from bit-level transmission toward a task-driven, ``understanding-before-transmission'' communication paradigm.

\subsubsection{JSCC}
Unlike the traditional paradigm that separates source and channel coding (SSCC), JSCC simultaneously exploits source statistics and channel conditions during encoding, enabling a more efficient reduction of redundancy while meeting target error-rate or distortion constraints. 
% JSCC adopts variable-length coding guided by feature importance, ensuring that critical semantic information is prioritized in both protection and transmission. This property makes it particularly suitable for scenarios with strict latency or bandwidth constraints. 
In the JSCC framework, a semantic extraction module first derives feature maps from raw data, achieving compression while preserving task- or reconstruction-relevant source information. The extracted features are then fed into a JSCC encoder that jointly performs compression and error-resilient mapping, where an entropy model is introduced to enable adaptive control of the coding length. Specifically, the entropy model (e.g., a learned probabilistic prior) estimates the probability distribution of features across spatial locations or channels and computes the corresponding information entropy or coding cost, thereby quantifying their semantic importance. Features with higher entropy (i.e., higher uncertainty and richer information content) are allocated longer codewords or more bits, whereas features with lower entropy are assigned shorter codewords and undergo stronger compression, realizing variable-length semantic source coding. Meanwhile, the importance estimates can also guide unequal error protection in channel coding or modulation resource allocation, enhancing the transmission reliability of critical features. Tailored to diverse semantic tasks and source modalities, customized semantic encoders and JSCC architectures can be deployed. The models are jointly trained via loss functions based on task-oriented metrics or reconstruction fidelity, thereby achieving adaptive and robust transmission under varying bandwidth and channel conditions.

Compared to legacy mobile systems, JSCC significantly mitigates the ``cliff effect'' in low-SNR or fast-fading channels and enables smooth performance degradation, thereby providing users with a more stable quality of experience (QoE)~\cite{xu2021wireless, tung2022deepwive, yao2022semantic, xiao2023wireless}. JSCC can be widely applied across the user plane, control plane, and data plane, as well as in application-specific scenarios and task-oriented services. To facilitate commercialization, the standard should specify baseline requirements for error-tolerance settings, variable code-length alignment, and E2E performance evaluation to keep different implementations consistent and interoperable.

\subsubsection{SemCom-based MA}
Current SemCom-based MA schemes can be classified into two main categories. The first category builds upon existing MA schemes, such as orthogonal frequency division multiple access (OFDMA), non-orthogonal multiple access (NOMA), and rate-splitting multiple access (RSMA)~\cite{gao2025rate}, by integrating semantic encoders and decoders to enable efficient compression and robust transmission. The second category explores novel MA techniques tailored for semantic data, with the most representative example being MDMA~\cite{zhang2023model}. The MDMA approach extends SemCom to multi-user scenarios by introducing a novel semantic domain beyond traditional resource dimensions such as time, frequency, and space. In MDMA, each transceiver pair employs a dedicated SemCom encoder-decoder model. Semantic features generated by different users occupy the same wireless resource blocks, enabling high resource reuse. These features are distinguished at the receiver using their respective SemCom models, allowing efficient and interference-resilient multi-user communication in a shared transmission environment.
Since the outputs of different encoding models remain approximately orthogonal in the semantic embedding space, this natural isolation lets the system extract the desired user content on demand while greatly improving spectral efficiency. 

\subsubsection{SemCom-based CSI Feedback}
Within a SemCom framework, CSI feedback is first transformed from raw time-frequency domain data into the spatial-frequency or delay-Doppler domain, followed by sparsification. A JSCC or JSCCM scheme is leveraged to deeply compress the sparse representation, significantly reducing uplink feedback overhead~\cite{xu2022deep, ren2025semcsinet}. Since CSI is updated periodically and exhibits temporal correlation, an additional long short-term memory (LSTM) network can model these time-domain dependencies, reducing redundancy while improving reconstruction accuracy. In high‑speed railway scenarios, where rapid Doppler shifts cause the channel to vary quickly, embedding Doppler features into the CSI semantic vector enables the encoder to learn and compensate for such effects at the semantic level, thereby lowering transmission loss and stabilizing the link. In 3GPP RAN1 and RAN2, research efforts for 6G standardization also focus on signaling procedures and technologies related to model LCM within CSI feedback, including data collection, model updating, and two-sided model alignment.

\subsubsection{Semantic KB Technologies}
The semantic KB can be constructed by various forms, including agents, datasets, vector codebooks, cross-modal features, and triple-based knowledge graphs. 
Semantic knowledge bases are most commonly instantiated as learnable semantic codebooks built upon Vector Quantization (VQ) and Variational Autoencoders (VAE). Such codebooks provide a principled way to discretize high-dimensional continuous semantic features and store them in a structured form. In this framework, a semantic extraction network first maps raw sources into a latent semantic space, after which VQ- or VAE-based mechanisms convert continuous embeddings into a finite set of discrete semantic symbols (i.e., codeword indices), yielding a shared codebook between the transmitter and receiver. VQ-based methods perform explicit discretization via nearest-neighbor matching, where each codeword represents a typical semantic prototype. In contrast, VAE-based methods learn a probabilistic latent distribution with prior constraints, capturing the statistical structure and uncertainty of semantic features. Acting as a shared semantic KB, the semantic codebook supplies essential prior information for the processes of semantic encoding, compression, importance assessment, and reconstruction. Consequently, it effectively curtails communication overhead while bolstering semantic fidelity and adaptability across varying tasks and channel states.

In the case of vector codebooks, the local semantic KB at the transmitter searches for the feature vector most similar to the content to be transmitted and sends only its index. The receiver then uses the same semantic KB to retrieve the corresponding feature vector and reconstruct the semantic feature map, significantly reducing overhead while maintaining service quality~\cite{wang2024unified, yi2023deep}. In the case of cross‑modal features, the semantic KB retrieves an associated text vector and fuses it with the image feature to achieve ``heterogeneous semantic complementation.'' Additionally, the KB should also support online learning and regional updates so that new scenario semantic features can be absorbed dynamically. The standards need to specify the KB format, synchronization mechanisms, index mapping, and security isolation rules to ensure consistency and interoperability across multi-vendor devices.

The SemCom architecture and its enabling technologies provide a solid foundation for 6G native-AI networks. However, it is essential for SemCom technologies to compatible with the existing protocol stack to maximize transmission performance without disrupting established procedures.

\section{Semantic Communication Compatibility Design}

\begin{figure*}[!t]
    \centering
    \includegraphics*[width=140mm]{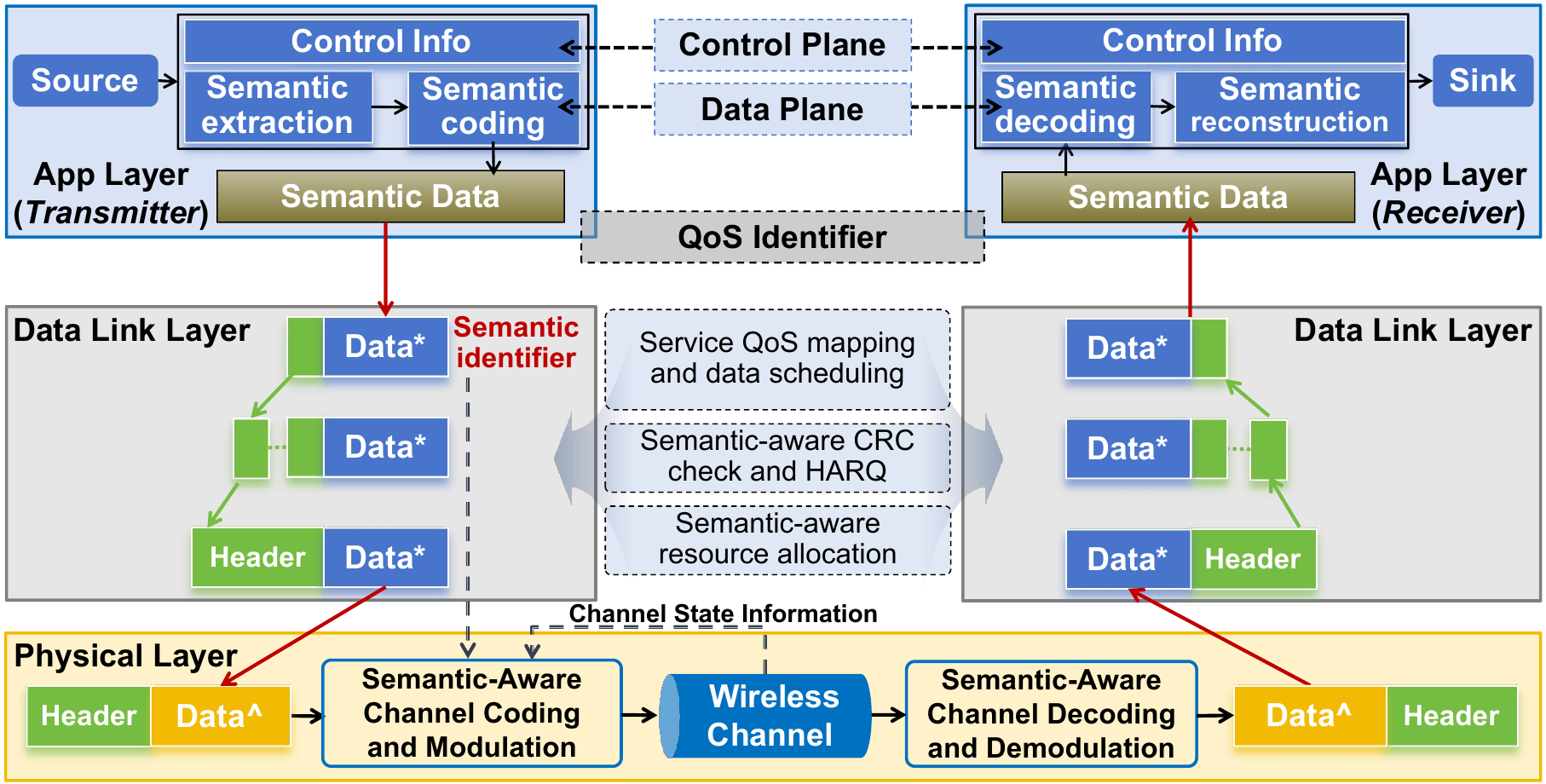}
    \caption{Compatibility design framework for semantic-traditional communication systems. The framework overlays new semantic-aware blocks onto the traditional mobile-network protocol stack to support seamless deployment of SemCom alongside existing procedures such as CRC, HARQ, and resource scheduling.}
    \label{fig:cross_layer}
\end{figure*}

The compatibility design of semantic-traditional communication system introduces semantic-aware processing modules while retaining the existing mobile communication protocol stack structure, enabling the integration of SemCom into conventional systems, as shown in Fig.~\ref{fig:cross_layer}. 
At the application layer, new modules for semantic extraction and semantic encoding/decoding are added to compress raw information into task-relevant semantic data groups. 
At the data link layer, semantic importance identifiers facilitate cross-layer mapping, enabling semantic-aware cyclic redundancy check (CRC), HARQ, and resource block allocation. 
At the physical layer, channel coding and modulation scheduling are adapted based on semantic importance to enhance the transmission reliability of critical semantic units. 
The overall architecture coordinates semantic feature transmission via QoS identifiers and control channels, allowing for flexible deployment and progressive evolution of SemCom without disrupting existing communication mechanisms.

\subsection{Physical Layer Design}

At the physical layer, SemCom introduces several key technologies to enhance the transmission efficiency and robustness of semantic information. First, semantic-aware channel coding and modulation techniques dynamically adjust coding rates and modulation schemes based on the importance of semantic units, ensuring higher reliability for critical content \cite{gao2024cross}. Second, power control and resource scheduling guided by semantic importance allow for efficient allocation of power and spectrum under resource-constrained conditions, reducing the risk of semantic distortion. Some studies have further explored E2E semantic encoders at the physical layer based on deep neural networks, which directly map high-level semantic goals into transmittable signal representations, thus breaking the limitations of traditional layered architectures \cite{ren2025semcsinet}. These physical layer innovations provide support for building semantic-aware and resilient 6G communication systems.

Furthermore, at the physical layer, differentiated processing of semantic features with varying importance can be integrated with technologies such as channel estimation, waveform design, and multi-antenna schemes to enable semantic-aware physical layer configuration. For example, in massive multiple-input multiple-output (MIMO) scenarios, semantic features can be efficiently mapped to precoding matrices to ensure the accurate recovery of task-relevant or information-critical semantic features.

\subsection{Network and Data Link Layer Design}

At the network and data link layers, SemCom incorporates a range of mechanisms to enable fine-grained management and reliable transmission of semantic information. First, the semantic feature-QoS priority mapping mechanism assigns different service levels to extracted semantic units based on their task relevance, facilitating semantic-driven logical channel mapping and scheduling control. Second, the semantic-level HARQ mechanism integrates semantic error sensitivity and task objectives into the retransmission strategy by introducing confidence-based semantic decision rules and differentiated retransmission thresholds, prioritizing retransmissions of semantic features most critical to task execution~\cite{gao2024cross}. Finally, semantic-level error control mechanisms, semantic-aware CRC, enable prioritized error correction for task-relevant semantic elements, thereby improving task success probability~\cite{gao2024cross}. Overall, these mechanisms significantly enhance the intelligence and adaptability of SemCom systems in multi-user and heterogeneous QoS environments, laying a technical foundation for building task-oriented and adaptive data link control frameworks.

\subsection{Control Plane Design}
To enhance control signaling in 6G systems, SemCom has been applied to both CSI feedback and HARQ-ACK/NACK transmission. For CSI, JSCC and JSCCM are implemented through a two-sided model at the network and users, enabling efficient compression, model alignment, and dynamic updates to reduce feedback overhead while maintaining accuracy~\cite{xu2022deep, ren2025semcsinet}. For HARQ-ACK/NACK signaling, AI-aided JSCCM replaces traditional repetition coding and fixed modulation by leveraging learned codebooks and content-aware power control. This approach exploits the inherent asymmetry and correlation in ACK/NACK bits, allocating resources based on semantic relevance and achieving significant gains in block error rate and SNR efficiency. Together, these techniques improve the efficiency and reliability of control plane, while maintaining compatibility with standardization frameworks through model deployment and updates. 
Additionally, effective deployment of SemCom-enabled CSI feedback requires comprehensive LCM, including performance evaluation, adaptive model switching, training data collection, and alignment between both network-device models. These capabilities are essential to ensure that the semantic representation of CSI remains accurate, up-to-date, and robust across dynamic mobile environments.

\subsection{Future Data Plane Design}
The concept of a unified data framework has been embraced by 3GPP SA2 to support data-centric and native-AI network architectures. This framework encompasses a wide range of data types, including wireless sensing data (e.g., obtained from integrated sensing and communication systems), AI/ML-related data (e.g., model parameters, training metadata, and inference results), network operation data (e.g., digital twin (DT) models, traffic profiles, and performance metrics), and heterogeneous IoT data. This massive volume of internal data imposes significant pressure on transmission and storage resources. SemCom alleviates this burden by enabling intent-aware and semantic-level information extraction, thereby facilitating precise and on-demand data transmission. SemCom enhances transmission efficiency in 6G networks by transmitting only task-relevant semantic features instead of raw data, thereby reducing redundancy and easing the burden on network resources. By enabling semantic-aware extraction, encoding, and forwarding, SemCom constructs an intelligent data plane that prioritizes meaning and task relevance, ultimately optimizing bandwidth utilization and improving the responsiveness and efficiency of native-AI 6G network functions.

% In digital twin use cases, SemCom serves as a key enabler for real-time, intelligent synchronization between physical and virtual network domains in native-AI 6G systems. By extracting and transmitting high-level, task-relevant features, such as traffic patterns, service intents, and QoS priorities, directly from heterogeneous edge data, SemCom avoids raw data overload and ensures accurate reconstruction of network states with minimal overhead. This meaning-aware, low-latency interaction enhances the digital twin’s ability to predict QoE risks, validate strategies through closed-loop feedback, and adapt dynamically to traffic changes, thereby reducing transmission burden, accelerating optimization cycles, and significantly improving end-to-end service quality and network autonomy.

\section{Case Study}

% We introduce a case study for CSI feedback, where the CSI‑feedback framework is implemented with a Transformer encoder-decoder backbone. The model is trained and evaluated on the publicly released CSI dataset from the mobile‑AI open platform\hyperlink{footnote5}{\footnotemark[5]}. Additionally, we apply implicit CSI feedback, and the over‑the‑air channel follows the 3GPP TR 38.901 clustered delay line (CDL) model, while the receiver employs least‑squares (LS) channel estimation to emulate the imperfect channel estimation conditions expected in practice. As shown in Fig.~\ref{fig:case study csi}, JSCCM achieves the highest SGCS across the entire SNR range thanks to the robustness of semantic model, it also achieves a higher compression ratio than the SSCC schemes at medium‑to‑high SNR, confirming its superior noise robustness and spectrum efficiency under standardized CDL fading.

\begin{figure}[t]
    \centering
    \includegraphics*[width=85mm]{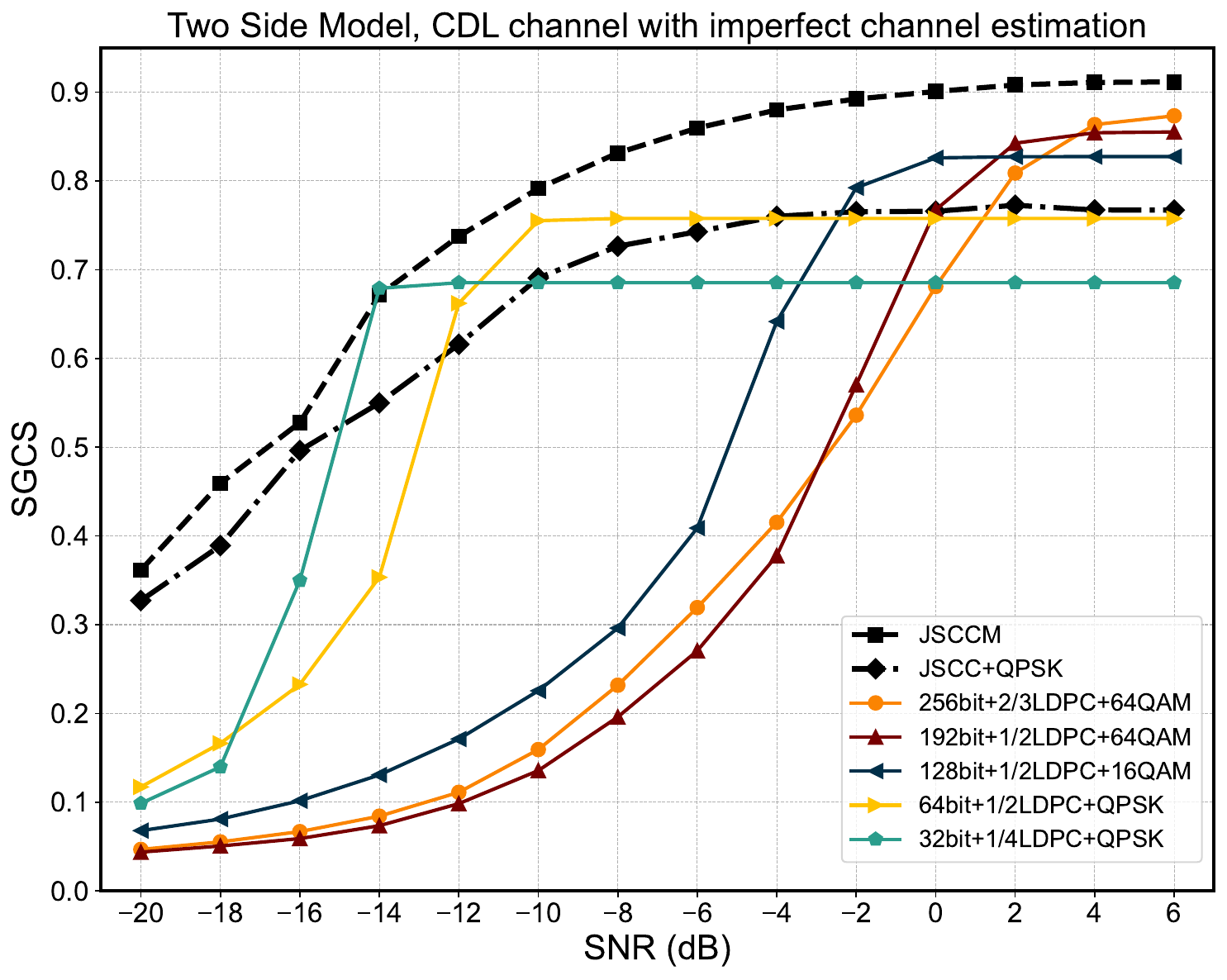}
    \caption{Squared generalized cosine similarity (SGCS) vs. SNR for the CSI feedback case study. ``JSCCM'' denotes the joint source-channel coding and modulation scheme. ``JSCC+QPSK'' applies deep JSCC for compression and quadrature phase shift keying (QPSK) for modulation. The remaining schemes represent SSCC: ``32bit+1/4LDPC+QPSK'' (and similar variants) indicates that an AI model performs source coding, producing a 32-bit representation that is then protected by a rate-1/4 LDPC code and modulated with QPSK. ``JSCC+QPSK'' and all SSCC baselines use 2-bit quantization so that each scheme transmits 128 symbols over the air.}
    \vspace{-4mm}
    \label{fig:case study csi}
\end{figure}

We introduce a CSI-feedback case study using the publicly released CSI dataset from the Mobile-AI open platform\hyperlink{footnote5}{\footnotemark[5]}. We assume implicit CSI feedback, the over-the-air link follows the 3GPP TR 38.901 clustered delay line (CDL) model, and the receiver uses least-squares (LS) channel estimation to emulate imperfect CSI. Key settings are carrier frequency $3.5 \ \text{GHz}$, $15 \ \text{kHz}$ subcarrier spacing with $64$ subcarriers and a single OFDM symbol per block, and delay spread $100 \ \text{ns}$. As shown in Fig.~\ref{fig:case study csi}, the encoders and decoders in all schemes utilize the two-layer Transformer as backbone and are trained E2E with the loss $1-\text{SGCS}$. Under the CDL channel, JSCCM attains the highest SGCS across the full SNR range, confirming superior noise robustness and transmission efficiency.

Additionally, SemCom has demonstrated effectiveness in immersive communication, intelligent internet of vehicles, multi-AI agent collaboration, satellite communications, and other representative scenarios \cite{guo2024survey, zhang2024intellicise, yang2022semantic, gao2025rate}, where both semantic-aware coding and knowledge-assisted reconstruction reduce resource consumption, while improving transmission performance.

\hypertarget{footnote5}{\footnotetext[5]{Dataset available at the Mobile-AI: \url{https://www.mobileai-dataset.cn/html/default/yingwen/DateSet/1589796123684470785.html?index=1}}}

\section{Way-forward of 6G Standardizations with Semantic communications}
\subsection{Unified Evaluation Metrics and Validation Framework}
The SemCom paradigm requires systematic scenario classification and performance characterization to support standardization across diverse service types, including control signaling, real-time media, multimodal sensing, and task-driven services. By mapping scenarios to tailored SemCom technologies, deployment priorities can be aligned with practical requirements.
To support fair comparison and verifiability across different implementations, a set of consistent, reproducible, and cross-layer evaluation metrics should be established, balancing semantic fidelity, task completion rate and accuracy, intent alignment, perceptual quality, and legacy indicators such as throughput and latency. These metrics are essential for the effective design and benchmarking of semantic-native networks. 

Additionally, scalable testbeds and simulation platforms shall be designed to emulate realistic communication environments, including noise, mobility, and interference, while preserving semantic fidelity across the E2E pipeline. Together with a unified evaluation framework, the unified validation framework are also essential for guiding both industry and academia toward the co-development of interoperable, verifiable, and practically deployable SemCom systems.
\vspace{-3mm}

\subsection{Foundational Architecture and Compatibility Design} 

A key step in advancing 6G standardization with SemCom is the establishment of a unified E2E architecture that supports semantic-level extraction and representation, semantic coding and modulation, and interaction across network layers. This architecture should clearly define core functional modules such as semantic encoder, decoder, inference, feedback, and QoS priority scheduling. it is essential to maintain compatibility with existing mobile communication systems to enable hybrid deployments and smooth integration. By extending from the control plane (e.g., CSI feedback) to the user and data planes, this architecture serves as the foundation for intelligent, task-aware, and native-AI communication systems in 6G.
\vspace{-3mm}
\subsection{Unified Platforms and Protocol Ecosystem}
The development of a unified simulation and experimental platform is essential to support standardized technology verification for SemCom. Such a platform should enable E2E evaluation across diverse scenarios, integrating semantic encoding, cross-layer transmission, and task-level performance assessment. By providing reproducible testing environments, standardized datasets, and configurable network modules, it ensures consistency and comparability of results across different implementations. This unified platform also facilitates collaborative validation among industry and academia, accelerating the standardization process and bridging the gap between theoretical research and practical deployment.

To enable scalable and trustworthy SemCom deployment in 6G native-AI systems, it is crucial to establish standardized, open-source models and reference implementations that ensure transparency, reproducibility, and interoperability. A modular and extensible standard should support diverse industries such as embodied intelligence and vehicular collaborative perception, through semantic-aware protocols, metadata formats, signaling extensions, model exchange, and legacy fallback solutions. Specifically, semantic-aware protocols refer to scheduling and transmission mechanisms that incorporate semantic importance when semantic traffic is transported through the network or when SemCom is integrated into existing communication systems.
Given the heterogeneity of AI models and KBs, unified application programming interfaces (APIs) and modular interfaces are essential. Achieving this vision requires close collaboration among standardization bodies (e.g., 3GPP, ITU, IEEE), industry, and academia to build a cohesive and trusted protocol ecosystem for future 6G networks.
% To enable scalable and trustworthy deployment of SemCom in 6G native-AI systems, it is essential to establish standardized, open-source models and reference implementations that ensure transparency, reproducibility, and cross-vendor interoperability. A modular, extensible, and cross-domain standard should support diverse vertical industries such as embodied intelligence, low-altitude unmanned aerial vehicles (UAVs), and cooperative vehicular perception by reflecting practical use cases through well-defined semantic-aware protocols and interfaces. This includes standardized semantic metadata formats, control plane signaling extensions, model exchange mechanisms, and fallback solutions for communication between semantic-enabled and legacy nodes. Given the heterogeneity of AI models and KB in SemCom systems, defining unified application programming interfaces (APIs) and modular interfaces is critical to ensure interoperability across vendors and domains. Achieving this vision will require close collaboration among standardization organizations (e.g., 3GPP, ITU, IEEE), industry players, and the research community to develop a cohesive and trusted protocol ecosystem that underpins future native-AI 6G networks.

\section{Conclusion}
In this paper, we have outlined a standardization roadmap for SemCom tailored to native-AI 6G systems. We reviewed representative application scenarios and proposed a unified evaluation framework that reflects the shift from bit-centric to meaning-aware metrics for different communication tasks. On the technical front, we identified several enabling technologies, such as JSCC, SemCom-based MA, SemCom-based CSI feedback, and semantic KB, as foundational candidates for standardization. To ensure backward compatibility with existing mobile networks, we introduced a compatibility design framework that supports SemCom as an attachable enhancement to legacy systems, offering a promising candidate for 6G evolution. Finally, we have discussed the key challenges ahead and presented a forward-looking vision for the standardization of SemCom in future native-AI 6G networks.

\section*{Acknowledgements}
This paper is supported by National Science and Technology Major Project of China on Mobile Information Networks under Grant 2024ZD1300700, in part by the National Natural Science Foundation of China No. 62401074, in part by the Beijing Natural Science Foundation No. L242012, in part by Young Elite Scientists Sponsorship Program of the Beijing High Innovation Plan, No. 20251037, and in part by the BUPT-CMCC Joint Institute. 

\footnotesize
% \bibliographystyle{IEEEtran}
% \bibliography{ref}

\section*{Biographies}

\vspace{3mm}
\noindent \textsc{Ping Zhang} (Fellow, IEEE) (pzhang@bupt.edu.cn)
is currently a professor of School of Information and Communication Engineering at Beijing University of Posts and Telecommunications, the director of State Key Laboratory of Networking and Switching Technology, a member of IMT-2020 (5G) Experts Panel, a member of Experts Panel for China’s 6G development. He served as Chief Scientist of National Basic Research Program (973 Program), an expert in Information Technology Division of National High-tech R\&D program (863 Program), and a member of Consultant Committee on International Cooperation of National Natural Science Foundation of China. His research interests mainly focus on wireless communication. He is an Academician of the Chinese Academy of Engineering (CAE).

\vspace{3mm}
\noindent \textsc{Xiaodong Xu} (Senior Member, IEEE) (xuxiaodong@bupt.edu.cn)
received his B.S degree in Information and Communication Engineering and Master’s Degree in Communication and Information System both from Shandong University in 2001 and 2004 separately. He received his Ph.D. degrees of Circuit and System in Beijing University of Posts and Telecommunications (BUPT) in 2007. He is currently a professor of BUPT and research fellow of Peng Cheng Laboratory. He has coauthored nine books and more than 120 journal and conference papers. He is also the inventor or co-inventor of 51 granted patents. His research interests cover semantic communications, intellicise communication system, moving networks, mobile edge computing and caching.

\vspace{3mm}
\noindent \textsc{Mengying Sun} (Member, IEEE) (smy\_bupt@bupt.edu.cn)
received the B.S. degree in communication engineering from Beijing University of Chemical Technology, Beijing, China, in 2016, and the Ph.D. degree in information and telecommunications engineering from Beijing University of Posts and Telecommunications (BUPT), Beijing, in 2022. From March 2021 to March 2022, she was a visiting Ph.D. student with the Department of Electrical and Computer Engineering, University of Waterloo, Waterloo, ON, Canada. She is currently an Assistant Professor with BUPT. Her research interests include mobile-edge computing, semantic communications, and AI-driven networking and communications.

\vspace{3mm}
\noindent \textsc{Haixiao Gao} (haixiao@bupt.edu.cn)
received the B.S. degree from the School of Electronics and Information Engineering, Tiangong University, Tianjin, China, in 2022. He is currently pursuing the Ph.D. degree with the School of Information and Communication Engineering, Beijing University of Posts and Telecommunications, Beijing, China. His research interests include semantic communication and deep learning.

\vspace{3mm}
\noindent \textsc{Nan Ma} (Member, IEEE) (manan@bupt.edu.cn)
received the B.S. and Ph.D. degrees from the Beijing University of Posts and Telecommunications (BUPT), China, in 2002
and 2007, respectively. He is currently a Professor with the School of Information and Communication Engineering, BUPT. His research interests include wireless communication testing theory and semantic communication.

\vspace{3mm}
\noindent \textsc{Xiaoyun Wang} (wangxiaoyun@chinamobile.com) 
received the B.S. and M.S. degrees in electrical engineering from the Beijing University of Posts and Telecommunications, Beijing, China, in 1991 and 2002, respectively. She is currently the Chief Scientist of China Mobile Communications Group Corporation. Her research interests include 6G network architecture, integrated architecture of computing and networking, and network intelligence.

\vspace{3mm}
\noindent \textsc{Ruichen Zhang} (Member, IEEE) (ruichen.zhang@ntu.edu.sg)
received the Ph.D. degree in Beijing Jiaotong University, Beijing, China. He is the postdoctoral research fellow in the College of Computing and Data Science, Nanyang Technological University, Singapore. His research interests include the Internet of Agents, LLM-empowered networking, reinforcement learning-enabled wireless communication, generative AI models, and heterogeneous networks. 

\vspace{3mm}
\noindent \textsc{Jiacheng Wang} (Member, IEEE) (jiacheng.wang@ntu.edu.sg)
earned his Ph.D. in Communication and Information Engineering from Chongqing University of Posts and Telecommunications, China, in 2022. He is now a postdoctoral researcher at the College of Computing and Data Science, Nanyang Technological University, Singapore, where his work focuses on integrated sensing and communications, generative artificial intelligence, and semantic communications.

\vspace{3mm}
\noindent \textsc{Dusit Niyato} (Fellow, IEEE) (dniyato@ntu.edu.sg)
is currently a professor in the College of Computing and Data Science, at Nanyang Technological University, Singapore. He received B.Eng. from King Mongkuts Institute of Technology Ladkrabang (KMITL), Thailand in 1999 and Ph.D. in Electrical and Computer Engineering from the University of Manitoba, Canada in 2008. His research interests are in the areas of Internet of Things (IoT), machine learning, and incentive mechanism design.

\end{document}